\documentclass[a4paper,11pt]{article}
\usepackage{jheppub}

\usepackage[T1]{fontenc}
\usepackage{CJK}
\usepackage{xcolor}
\usepackage{amsfonts}
\usepackage{epstopdf}
\usepackage{wrapfig} % to wrap figure with
\usepackage{subfigure}
\usepackage{graphicx}  % Include figure files
\usepackage{dcolumn}   % Align table columns
\usepackage{bm}
\usepackage{float}

\newcommand{\be}{\begin{equation}}
\newcommand{\ee}{\end{equation}}
 \newcommand{\bea}{\begin{eqnarray}}
 \newcommand{\ena}{\end{eqnarray}}

\newcommand{\mbf}{\bm}

\newcommand{\rvec}{\rightarrow}
\newcommand{\bra}{\langle}
\newcommand{\ket}{\rangle}

\newcommand{\smalleq}{\leqslant}
\newcommand{\linear}{\propto}

\newcommand{\notice}{\equiv}

\newcommand{\exd}{\mathrm{d}}
\newcommand{\grad}{\nabla}
\newcommand{\pd}{\partial}

\title{Holographic Einstein Ring of a Charged AdS Black Hole}
\author[1]{Yuxuan Liu}
\author[2]{Qian Chen}

\author[3,4]{Xiao-Xiong Zeng}
\author[5]{Hongbao Zhang}
\author[5]{Wenliang Zhang}

\affiliation[1]{Kavli Institute for Theoretical Sciences (KITS), University of Chinese Academy of Sciences, Beijing 100190, China}
\affiliation[2]{School of Physics, University of Chinese Academy of Sciences, Beijing 100049, China}
\affiliation[3]{State Key Laboratory of Mountain Bridge and Tunnel Engineering, Chongqing Jiaotong University, Chongqing 400074, China}
\affiliation[4]{Department of Mechanics, Chongqing Jiaotong University, Chongqing 400074, China}
\affiliation[5]{Department of Physics, Beijing Normal University, Beijing 100875, China}
\emailAdd{liuyuxuan@ucas.ac.cn}
\emailAdd{chenqian192@mails.ucas.ac.cn}
\emailAdd{xxzengphysics@163.com}
\emailAdd{hongbaozhang@bnu.edu.cn}
\emailAdd{201921140020@mail.bnu.edu.cn}

\abstract{\\   Taking into account that the real quantum materials are engineered generically at a finite chemical potential, we investigate the Einstein ring structure for the lensed response of the complex scalar field as a probe wave on the charged AdS black hole in the context of AdS/CFT. On the one hand, we find that the resulting Einstein ring radius has no variation with the chemical potential, which is similar to the behavior for the weakly interacting quantum system. On the other hand, not only can such a ring exist well within the screen, but also the temperature dependence of its radius exhibits a distinct feature in the sense that it displays an appreciable increase at low temperatures while the ring keeps unchanged right at the edge of the screen for the weakly interacting system. Note that such a Einstein ring emerges in the large frequencies and can be well captured by the photon sphere away from the black hole horizon in the geometric optics approximation, thus such a distinct feature may be regarded as a universal behavior associated with the high energy modes of the strongly coupled system which has a gravity dual. }

\begin{document}
\maketitle
\flushbottom

\section{Introduction}\label{intro}
Since its advent\cite{Maldacena,GKP,Witten}, AdS/CFT correspondence has emerged as a powerful tool in helping understand the strongly correlated dynamics of quantum many body systems\cite{Hartnoll,HLS,LS,BGHLL}, where some universal low energy behaviors of the boundary system are obtained often by the near horizon geometry of the bulk black hole\cite{viscosity,viscosity2,SS,MSS,complexity,complexity2,BDS,HHM,BDGL,AAGM,Grozdanov}. It is interesting to ask whether there also exist some universal features associated with the high energy collective excitations of the strongly coupled systems. As shown more recently along this line for a finite temperature strongly coupled system which has a gravity dual, the lensed response displays a universal Einstein ring structure due to the existence of the photon sphere outside of the bulk black hole if one puts a monochromatic axi-symmetric Gaussian source in the vicinity of the south pole of the $2$-sphere on which the system lives\cite{HKM,HKM2}. Furthermore, as opposed to the weakly interacting quantum system, the size of the observed Einstein ring on the boundary varies with the photon sphere in the bulk. Thus it is reasonable to suspect that such a photon sphere induced Einstein ring structure is supposed be ubiquitous in any holographic quantum matter. Put it another way, the appearance of such a Einstein ring structure in some quantum matter may be used as a strong signal for the existence of its gravity dual. But nevertheless, since the photon sphere varies according to the specific bulk dual geometry under consideration, the detailed behavior of the Einstein ring structure is also expected to vary. Therefore one is tempted to investigate the behavior of the lensed response for a variety of holographic quantum materials. Note that the Schwarzschild-AdS black hole considered in \cite{HKM,HKM2} corresponds only to the finite temperature boundary system at some fixed density of quantum particles commensurate with an underlying lattice\cite{HLS}. However, the majority of quantum materials are prepared not at such a special density. Rather, they are engineered at generic densities in a phase diagram in which the density, or equivalently the chemical potential can be controlled in a continuous manner. To describe them by holography, one is required to turn on the bulk electromagnetic field. Accordingly, the bulk black holes will be charged. The simplest holographic theory with the required ingredients is the Einstein-Maxwell-AdS theory, and its charged black hole solution is well known as the Reissner-Nordstrom-AdS black hole.

The purpose of this paper is to explore the characteristic behavior of the lensed response of the charged probe operator on top of the aforementioned holographic matter. As such, we are also required to introduce an additional bulk probe complex scalar field as the holographic dual to the charged probe operator. With this, not only shall we investigate how the temperature affects the lensed response, but also examine the effect of the chemical potential on the resulting Einstein ring.

The structure of this paper is organized as follows. In the next section, we shall introduce the retarded Green function to relate the response with the source in the linear response theory and the optical apparatus to image the lensed response function. For the later comparison, the corresponding ring formation for the monochromatic axi-symmetric source Gaussian source peaked on the south pole is also presented for a weakly interacting thermal system at finite chemical potential. In Section \ref{background}, we introduce our holographic model with the charged scalar field as a probe field propagating on top of the Reissner-Nordstrom-AdS background and present how to calculate the resulting response function by the wave dynamics in the bulk using the state of the art numerics. Taking into account that the sharp ring is expected to occur in the sufficiently large frequencies, we also work out the geometric optics approximation of the wave dynamics as well as the Hamilton-Jacobi formalism to solve the trajectories, where the generalized photon sphere is explained to be responsible for the radius of the formed Einstein ring. With the above preparation, we present our main numerical results in Section \ref{main}, where the variations of the Einstein ring with respect to the temperature and chemical potential are especially highlighted. We conclude our paper with some future directions worthy of further investigation in the last section. We relegate the calculation of the retarded Green function for the thermal quantum field theory at finite chemical potential and the pseudo-spetral method associated with the Chebyshev polynomials into Appendix \ref{AA} and Appendix \ref{spectral}, respectively.

\section{Retarded Green function, optical apparatus, and ring formation}\label{green}
%From Eq.(\ref{lwave}) we know that the Einstein ring can be calculated by the response function, while the response function is related to Green function, thus it is interesting to check whether the Einstein ring can be given by the Green function locating on boundary only.
For the equilibrium quantum system living on the sphere, the linear response function is relate to the source by the retarded Green function $G(t,\theta,\varphi; t',\theta',\varphi')$, which, due to the time translation symmetry and spatial rotation symmetry, can be expressed as
\begin{equation}
    G=\sum_{l=0}^\infty\sum_{m=-l}^{l}\int \frac{\exd\omega}{2\pi}e^{-i\omega(t-t')}Y_{lm}(\theta,\varphi)Y^*_{lm}(\theta',\varphi')G_{lm}(\omega).
\end{equation}
As in \cite{HKM,HKM2}, we choose the monochromatic and axi-symmetric Gaussian wave packet centered on the south pole $\theta_0=\pi$ as the source
\begin{equation} \label{source}
	J_\mathcal{O}(v,\theta)=e^{-i\omega v}\frac{1}{2\pi \sigma^2}\exp\left[-\frac{(\pi-\theta)^2}{2\sigma^2}\right]=e^{-i\omega v}\sum_{l=0}^\infty {c}_{l0}Y_{l0}(\theta),
\end{equation}
where the wave packet size is taken to be $\sigma\ll\pi$, and the coefficients of the spherical harmonics $Y_{l0}(\theta)$ can be calculated out as
\begin{equation}\label{coe}
 	c_{l0}=(-1)^l\sqrt{\frac{l+1/2}{2\pi}}\exp\left[-\frac12 (l+1/2)^2\sigma^2\right].
 \end{equation}
Accordingly, the corresponding response function is given by
\begin{equation}\label{pole}
    \bra\mathcal{O}\ket_{J_\mathcal{O}}=\int_0^{2\pi} \exd\varphi'\int_0^\pi \exd\theta'\sin\theta' G(t,\theta,\varphi;t',\theta',\varphi')J_\mathcal{O}(t',\theta'\varphi')=\sum_{l=0}^\infty e^{-i\omega t}G_{l0}(\omega)c_{l0}Y_{l0}(\theta).
\end{equation}
Whence we know that the dominant contribution to the response function is given by the mode $l$ for which the pole of $G_{l0}$ is the closest to the given frequency $\omega$.
\begin{figure}
	\centering
	%\begin{minipage}[t]{0.48\textwidth}
	%\centering
	\includegraphics[height=2.5in]{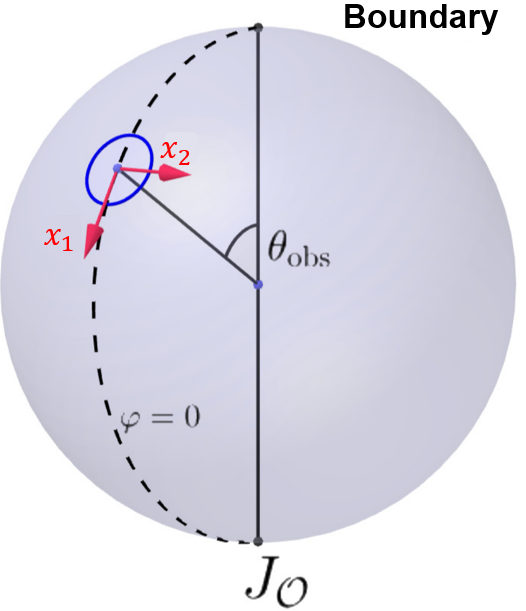}
	\caption{The observation region is surrounded by the blue circle on the boundary unit sphere.}\label{thetaOBS}
%	\end{minipage}
	\end{figure}
	
To observe the above response function, let us introduce the optical apparatus as follows.  We first suppose that the region we choose to observe is surrounded by a small circle with the center located at $(\theta_{\text{obs}},0)$ on the  unit sphere, which is demonstrated in Figure \ref{thetaOBS}. By rotating the original spherical coordinate system $\{\theta,\phi\}$ to a new one $\{\theta',\varphi'\}$ in such a way that
\begin{equation}
	\sin\theta'\cos\varphi'+i\cos\theta'=e^{i\theta_{\text{obs}}}(\sin\theta\cos\varphi+i\cos\theta),
\end{equation}
we have $(\theta'=0,\varphi'=0)$ corresponds to the center of the observation region. Furthermore, we would like to introduce a Cartesian coordinate system $\{x_1,x_2,x_3\}$ such that $(x_1,x_2)=(\theta'\cos\varphi',\theta'\sin\varphi')$ in the observation region.
	\begin{figure}
	    \centering
	%\begin{minipage}[t]{0.48\textwidth}
%	\centering
	\includegraphics[height=2.5in]{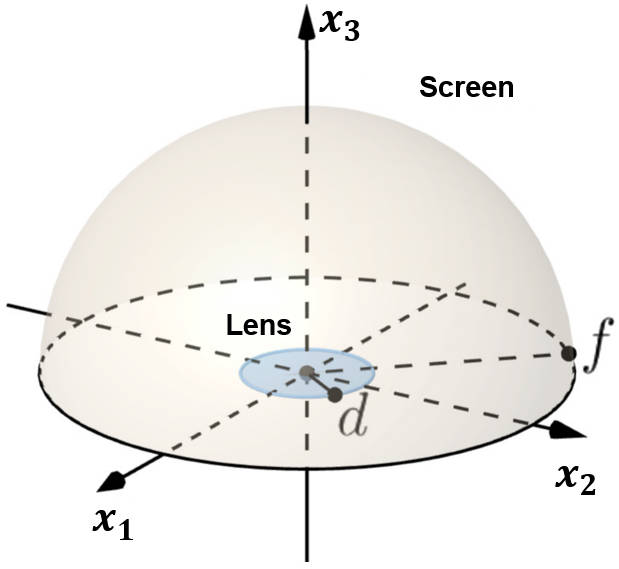}
	\caption{The convex lens of radius $d$ is put onto the observation region and the screen is adjusted at the focus of the convex lens.}\label{telescope}
%	\end{minipage}
\end{figure}
Next we introduce the imaging system, which consists of a convex lens and a spherical screen. The role of the convex lens is to transmit a wave $\Psi(\vec{x})$ into the following form
\begin{equation}
	\Psi_T(\vec x)=e^{-i\tilde{\omega}\frac{|\vec x|^2}{2f}}\Psi(\vec x),
\end{equation}
where $\tilde{\omega}=\omega+\mu$ with $\mu$ the chemical potential for the system and $f$ is the focus of the convex lens. As illustrated in Figure \ref{telescope}, with the convex lens put onto the above observation region and the spherical screen adjusted such that the points on the screen satisfies ${x_{S1}}^2+{x_{S2}}^2+{x_{S3}}^2=f^2$, the wave function recorded on the screen is given by
\begin{equation}\label{lensTranslation}
	\Psi_S(\vec x_S)=\int_{|\vec x|\smalleq d}\exd^2 x\Psi_T(\vec x)e^{i\tilde{\omega} D}
	\linear \int_{|\vec x|\smalleq d}\exd^2 x\Psi(\vec x) e^{-i\frac{\tilde{\omega}}{f}\vec x\cdot\vec x_S}= \int \exd^2 x\Psi(\vec x) w(\vec x) e^{-i\frac{\tilde{\omega}}{f}\vec x\cdot\vec x_S},
\end{equation}
where the integral is performed over the convex lens of radius $d$, $D$ is the propagating distance from the lens point $(x_1,x_2,0)$ to the screen point $(x_{S1},x_{S2},x_{S3})$, and $w(\vec x)$ is the window function, defined as
\begin{equation}
	w(\vec x):=\begin{cases}
		1,\quad 0\smalleq|\vec x|\smalleq d,\\
		0,\quad |\vec x|>d.
	\end{cases}
\end{equation}
According to the last expression of Eq. (\ref{lensTranslation}), the image will be formed at $\vec{x}_S=f \hat{n}$ on the screen for the incident wave $\Psi(\vec{x})\linear e^{i\tilde{\omega}\hat{n}\cdot \vec{x}}$ with $\hat{n}$ the normalized propagating direction vector. Such an observation justifies the familiar role of the convex lens.

\begin{figure}
	\centering
	\includegraphics[height=3in]{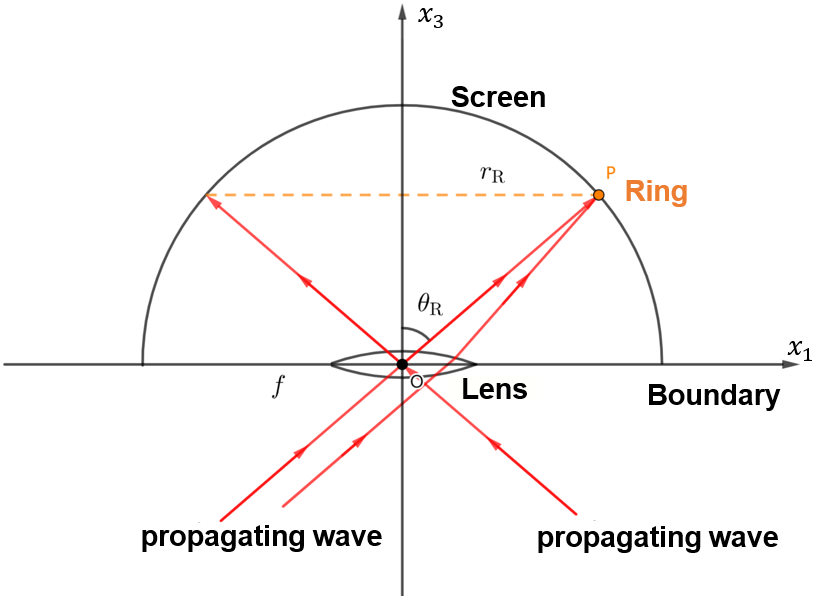}
	\caption{The relation between $\theta_\text{R}$ and $r_\text{R}$.}\label{thetaR}
\end{figure}

With the above apparatus, one can plot $|\Psi_S|^2$ as the image of the lensed response with $\Psi=\bra\mathcal O\ket_{J_\mathcal{O}}$. On physical grounds, we expect to see a sharp image when the frequency $\omega$ is sufficiently large. In particular, as detailed in Appendix \ref{AA} for a weakly interacting quantum field theory at finite temperature and finite chemical potential, the retarded Green function takes the following form
\begin{equation}
        G_{lm}(\omega)=\frac{1}{\tilde{\omega}^2-l(l+1)-m_T^2},
\end{equation}
where the thermal mass is given by $m_T^2=m^2+\frac{\lambda}{2} h(T,\mu)$ to the one-loop level with the typical variation of $h$ demonstrated in Appendix \ref{AA}. Thus the dominant contribution to the response function comes from the large $l$ mode satisfying
\begin{equation}
    l(l+1)\approx \tilde{\omega}^2-m_T^2.
\end{equation}
Note that in the large $l$ limit $Y_{l0}(\theta)$ can be approximated as the superposition of $e^{i l\theta}$ and $e^{-i l \theta}$. Thus as illustrated in Figure \ref{thetaR}, a ring will be formed on the screen with the angle of the straight line between the center of the lens and the point on the ring to the optical axis of the lens given by
\begin{equation}
    \sin\theta_R\approx\frac{l}{\tilde{\omega}}\approx 1-\frac{m_T^2}{2\tilde{\omega}^2},
\end{equation}
which amounts to saying
\begin{equation}
   \frac{ r_R}{f}\approx  1-\frac{m_T^2}{2\tilde{\omega}^2}.
\end{equation}
It is obvious that both the temperature and chemical potential dependence of the ring radius $r_R$ are suppressed in the large frequency limit. Accordingly, the ring keeps unchanged right at the edge of the screen, as opposed to the strongly coupled system which has a gravity dual, where the pole is given by the quasi-normal mode and captured essentially by the photon sphere in the large frequency limit.

\section{Holographic setup, wave dynamics, and geometric optics approximation}\label{background}

Let us start with the following bulk action
\begin{equation}
	I_{\text{bulk}}=\int\exd^4 x\sqrt{-g}\left[R+\frac{6}{{L_{\text{AdS}}}^2}-\frac14F^{ab}F_{ab}-|D\Phi|^2-M^2|\Phi|^2 \right],
\end{equation}
where $R$ is the Ricci scalar, $F=\exd A$ with $A$ the electromagnetic 4-potential, $D_a\notice \grad_a-ieA_a$ is the covariant derivative operator, and $\Phi$ is a complex scalar field with $e$ its electric charge and $M$ its mass. Below we shall set the AdS radius $L_{\text{AdS}}=1$ for simplicity.

Associated with the above action, we like to consider the background solutions with the following ansatz
\begin{eqnarray}
	&\exd s^2=-\hat{f}(r)e^{-\hat{\chi}(r)}\exd t^2+\frac{1}{\hat{f}(r)}\exd r^2+r^2\exd \Omega^2=\frac1{z^2}\left[-f(z)e^{-\chi(z)}\exd t^2+\frac{\exd z^2}{f(z)}+\exd\Omega^2\right],\\
	&A_a=-\hat{A}(r)(\exd t)_a=-A(z)(\exd t)_a,\quad \Phi=0
\end{eqnarray} \label{metric}
in the coordinate systems $\{t,r,\theta,\varphi\}$ and $\{t,z,\theta,\varphi\}$ with $\exd\Omega^2\notice\exd\theta^2+\sin\theta^2\exd\varphi^2$ the metric of the spatial $2$-sphere. These two coordinates are related by $z= r^{-1}$, whereby we have $\hat{f}(r)=z^{-2}f(z)$. $z=0$ corresponds to the AdS boundary where the dual quantum system lives. In particular, below we shall focus solely on the Reissner-Nordstrom-AdS (RN-AdS) black hole, i.e.,
\begin{align}
	f(z)&=\frac14\rho^2z^4+z^2+1-\left(\frac14\rho^2z_h^4+z_h^2+1\right) \frac{z^3}{z_h^3},\label{rnads1}\\
	A(z)&=\rho z-\rho z_h,\label{rnads2}\\
	\chi(z)&=0\label{rnads3},
\end{align}
where $\rho$ is the charge parameter of black hole, and $z_h$ denotes the location of black hole event horizon. $\rho=0$ corresponds to the Schwarzschild-AdS black hole. Furthermore, the pure-AdS spacetime can be obtained by taking the limit $z_h\rvec \infty$. By holography, $\mu=\rho z_h$ is interpreted as the chemical potential of the boundary system while the temperature of the boundary system is given by the Hawking temperature
\begin{equation}
    T=\frac{3+z_h^2-\frac{1}{4}\rho^2z_h^4}{4\pi z_h}.
\end{equation}
Note that different from \cite{HKM,HKM2}, where both the small and large black Schwarzschild black holes are considered no matter whether they are thermodynamic stable, we shall restrict ourselves to the regime where the above RN-AdS solution is thermodynamic stable in the grand canonical ensemble. In particular, when there are two black hole solutions at a given temperature and chemical potential, we only take the large black hole as our background\cite{CEJM}.

Next we take the complex scalar field as a probe field in the above RN-AdS background. The corresponding dynamics is governed by the Klein-Gordon equation
\begin{equation} \label{EOM}
	D_a D^a\Phi-M^2\Phi=0.
\end{equation}

To solve it by numerics in a more convenient manner, we prefer going to the ingoing Eddington coordinate, i.e.,
\begin{equation}
	v\notice t+z_*=t-\int \frac{e^{\chi(z)/2}}{f(z)}\exd z.
\end{equation}
As a result, the non-vanishing bulk background fields are transformed into the following smooth form
\begin{eqnarray}
	&\exd s^2=\frac{1}{z^2}\left[-f(z)e^{-\chi(z)}\exd v^2-2 e^{-\chi(z)/2}\exd z\exd v+\exd\Omega^2\right], \label{newmetric1}\\
	&A_a=-A(z)(\exd v)_a,
\end{eqnarray}
where the gauge transformation is also applied to the electromagnetic 4-potential. In what follows, we shall take $e=1$ and $M^2=-2$ for definiteness.  With $\Phi=z\phi$, the asymptotic behaviour of $\phi$ near the AdS boundary can be expressed as
\begin{equation}\label{vzbehaviour}
	\phi(v,z,\theta,\varphi)=J_\mathcal{O}(v,\theta,\varphi)
	+ \bra\mathcal{O}\ket z+O(z^2).
\end{equation}
By the holographic dictionary, $J_\mathcal{O}$ is interpreted as the source for the boundary field theory, and the corresponding expectation value of the dual operator, namely the response function, is given by
\begin{align}\label{res1}
	\bra\mathcal{O}\ket_{J_\mathcal{O}}=\bra\mathcal{O}\ket-(\pd_v-i\mu)J_\mathcal{O},
\end{align}
where $\bra\mathcal{O}\ket$ corresponds obviously to the expectation value of the dual operator with the source turned off.

With the source given by Eq. (\ref{source}), the corresponding bulk solution takes the following form
\begin{equation}\label{fieldDecomp}
	\phi(v,z,\theta)=e^{-i\omega v}\sum_{l=0}^\infty c_{l0}Z_l(z)Y_{l0}(\theta),
\end{equation}
where $Z_l$ satisfies the equation of motion
%\begin{align}\nonumber
%	  0=&z^{2} f{\left(z \right)}  {Z}_{l}''+ \left(z^{2}   f'{\left(z \right)} + 2 i %w z^{2}  - 2 i e z^{2} A{\left(z \right)} \right) {Z}_{l}'\\
 %     & + \left(z   f'{\left(z \right)} - 2 f{\left(z \right)} - i e z^{2}  %A'{\left(z \right)} - l(l+1) z^{2}  + 2 \right){Z}_{l},\label{zpart}
%\end{align}
\begin{align}
z^2fZ_l''+z^2[f'+2i(\omega-eA)]Z_l'+[(2-2f)+zf'-z^2(ieA'+l(l+1))]Z_l=0,\label{zpart}
\end{align}
and its asymptotic behaviour near the AdS boundary goes like
\begin{equation}\label{expansion}
 	 Z_l=1+\bra\mathcal{O}\ket_l z+O(z^2).
\end{equation}
Similarly, the resulting response $\bra\mathcal{O}\ket_{J_\mathcal{O}}$ can be expressed as
\begin{equation}\label{resDecomp}
	\bra\mathcal{O}\ket_{J_\mathcal{O}}=e^{-i\omega v}\sum_{l=0}^\infty c_{l0}\bra\mathcal{O}\ket_{J_\mathcal{O} l} Y_{l0}(\theta)
\end{equation}
with
\begin{align}
	\bra\mathcal{O}\ket_{J_\mathcal{O} l}=\bra\mathcal{O}\ket_l+i\tilde{\omega}.
\end{align}
\begin{figure}
	\centering
	\includegraphics[height=2in]{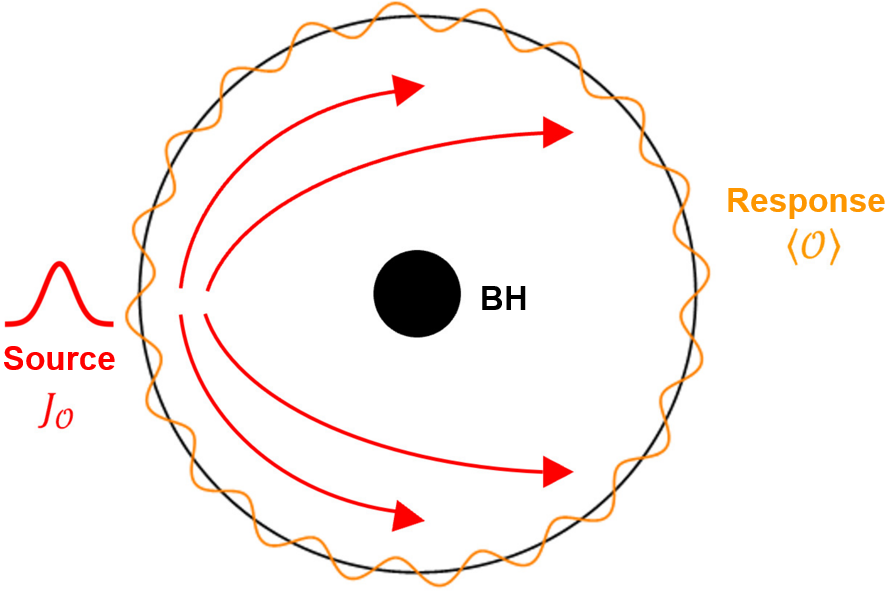}
	\caption{The monochromatic Gaussian source $J_\mathcal{O}$  is located at a point on the AdS boundary, and its response $\bra\mathcal{O}\ket$ is observed at another point on the boundary far away from the source.  }\label{sourceresponse}
\end{figure}
The key task is to solve the radial equation Eq. (\ref{zpart}) with
the following boundary condition
\begin{equation}
	Z_l(0)=1
\end{equation}
at the AdS boundary and the regular boundary condition on the black hole event horizon, which is the very advantage of the ingoing Eddington coordinate over the Schwarzschild coordinate used in \cite{HKM,HKM2}. On the other hand, we employ the pseudo-spectral method to obtain the corresponding numerical solution for $Z_l$ and extract $\bra\mathcal{O}\ket_l$, which turns out to be much more efficient and much more precise than the strategy adopted in \cite{HKM,HKM2}. With the extracted $\bra\mathcal{O}\ket_l$, the total response can be obtained by Eq. (\ref{resDecomp}). As illustrated in Figure \ref{sourceresponse}, the total response we consider below is far away from where the source is located, so we shall neglect the second term associated with the source in Eq. (\ref{res1}), which amounts to saying that $\bra\mathcal{O}\ket_{J_\mathcal{O}l}$ in Eq. (\ref{resDecomp}) can be replaced by $\bra\mathcal{O}\ket_l$. In addition, it follows from Eq. (\ref{coe}) that $c_{l0}$ decreases exponentially with the increase of $l$, so we shall truncate Eq. (\ref{resDecomp}) to the summation over $0\smalleq l\smalleq 200$.

\begin{figure}
	\centering
	\includegraphics[height=2in]{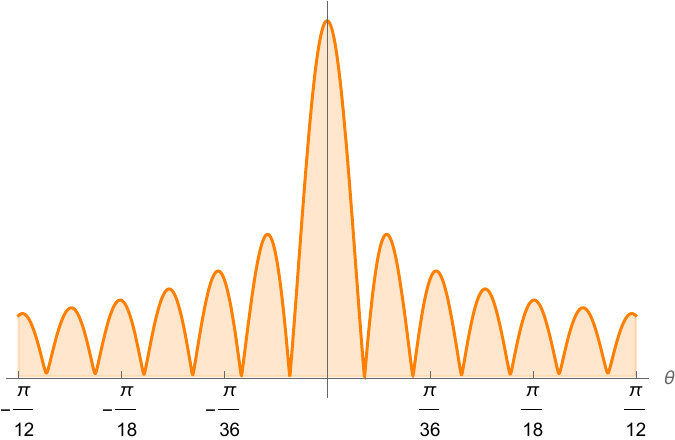}
	\caption{The amplitude of $|\bra \mathcal{O}\ket|$ around the north pole at $T=1/\pi$ and $\mu=2$ for $\omega=75$.}\label{resp-amp}
\end{figure}

\begin{figure}
    \centering
    \subfigure[$\omega=10$]{
        \includegraphics[width=1.3in]{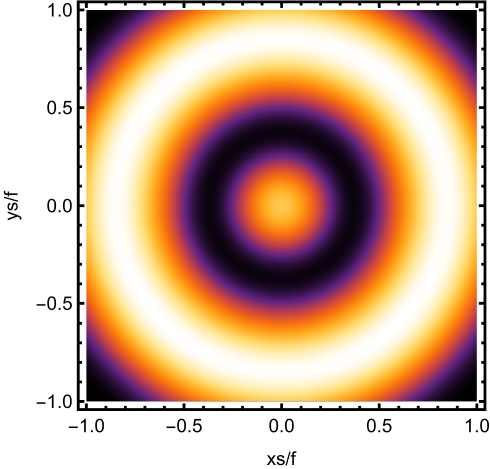}
    }
	\subfigure[$\omega=35$]{
        \includegraphics[width=1.3in]{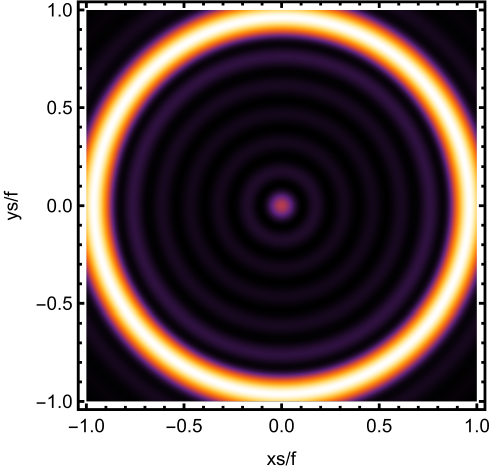}
    }
    \subfigure[$\omega=75$]{
        \includegraphics[width=1.3in]{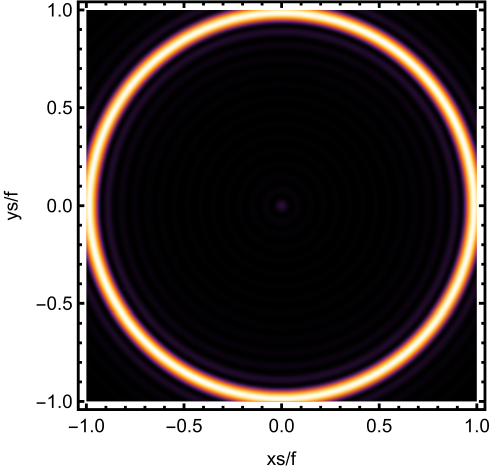}
    }
    \caption{The image of the lensed response on the screen at $T=1/\pi$ and $\mu=2$ for different frequencies, where $\frac{r_R}{f}$ of the sharp Einstein ring for $\omega=75$ is around $0.984$, in good agreement with $0.996$ given by the geometric optics approximation.}\label{sharpimage}
\end{figure}

As a demonstration, we plot a typical profile of the total response in Figure \ref{resp-amp}, where the apparent interference pattern arises from the diffraction of our scalar field off the black hole.
%In the next section, we shall transform the resulting response into the image on the screen by a convex lens and investigate the pattern of t
With the optical apparatus introduced in the previous section, one can obtain the image of the lensed response at the north pole. As demonstrated in Figure \ref{sharpimage} with $\sigma=0.05$ for the source and $d=0.6$ for the convex lens, the higher the frequency becomes, the sharper the resulting ring becomes. This is reasonable because it is expected that the image can be well captured by the geometric optics approximation in the high frequency limit. To be more specific, suppose that $\Phi=\mathcal{A}e^{iS}$ with the amplitude $\mathcal{A}$ slowly varying while the phase $S$ rapidly varying in the background, then the geometric optics approximation of  our wave equation (\ref{EOM}) gives rise to
\begin{equation}
g^{\mu\nu}U_\mu U_\nu=-M^2, \quad U^\mu\nabla_\mu\ln (\mathcal{A}^2)+\nabla^\mu U_\mu=0
\end{equation}
with $U_\mu=\partial_\mu S-eA_\mu$ interpreted as the four velocity of the corresponding trajectories. It is noteworthy that the phase $S$ can also be understood as the following on-shell action
\begin{equation}
S=\int \exd\eta (\frac{1}{2}g_{\mu\nu}\frac{\exd x^\mu}{\exd\eta}\frac{\exd x^\nu}{\exd\eta}+eA_\mu\frac{\exd x^\mu}{\exd\eta})
\end{equation}
of the special solution to the Hamilton-Jacobi equation
\begin{equation}\label{HJ}
\partial_\mu S=g_{\mu\nu}\frac{\exd x^\nu}{\exd\eta}+eA_\mu,\quad \frac{\partial S}{\partial\eta}+\frac{1}{2}g_{\mu\nu}\frac{\exd x^\mu}{\exd\eta}\frac{\exd x^\nu}{\exd\eta}=0
\end{equation}
by taking $\frac{\partial S}{\partial \eta}=\frac{1}{2}M^2$ at the end of the day. For our purpose, we like to go back to the coordinate system $(t,r,\theta,\varphi)$ and consider the trajectories with $\varphi$ fixed. Accordingly, $S$ does not depend on $\varphi$, which reduces Eq. (\ref{HJ}) to
\begin{equation}
    2\frac{\partial S}{\partial \eta}-\frac{1}{\hat{f}}(\partial_t S+eA)^2+\hat{f}(\partial_r S)^2+\frac{1}{r^2}(\partial_\theta S)^2=0.
\end{equation}
Note that both $\hat{f}$ and $A$ depend solely on $r$. Thus the corresponding general solution can be obtained readily by separation of variables as
\begin{equation}\label{phase}
    S=\frac{1}{2}\mathcal{N}\eta-\omega t+L\theta+\int^r \frac{\exd r}{\hat{f}}\sqrt{\mathcal{R}}
\end{equation}
with $\mathcal{R}=(\omega-eA)^2-\hat{f}(\frac{L^2}{r^2}+\mathcal{N})$.
 The trajectory can be further obtained by letting the partial derivative of $S$ with respect to $\mathcal{N}$, $\omega$, and $L$ equal to constants as follows
 \begin{eqnarray}\label{ray}
     \eta-\int \exd r\frac{1}{\sqrt{\mathcal{R}}}&=&\text{const.},\nonumber\\
     -t+\int \frac{\exd r}{\hat{f}}\frac{\omega-eA}{\sqrt{\mathcal{R}}}&=&\text{const.},\nonumber\\
     \theta - \int\frac{\exd r}{r^2}\frac{L}{\sqrt{\mathcal{R}}}&=&\text{const.}.
 \end{eqnarray}
As mentioned before, the final solution is given by taking $\mathcal{N}=M^2$ in Eq. (\ref{phase}) and Eq. (\ref{ray}).
\begin{figure}
	\centering
	\includegraphics[height=3in]{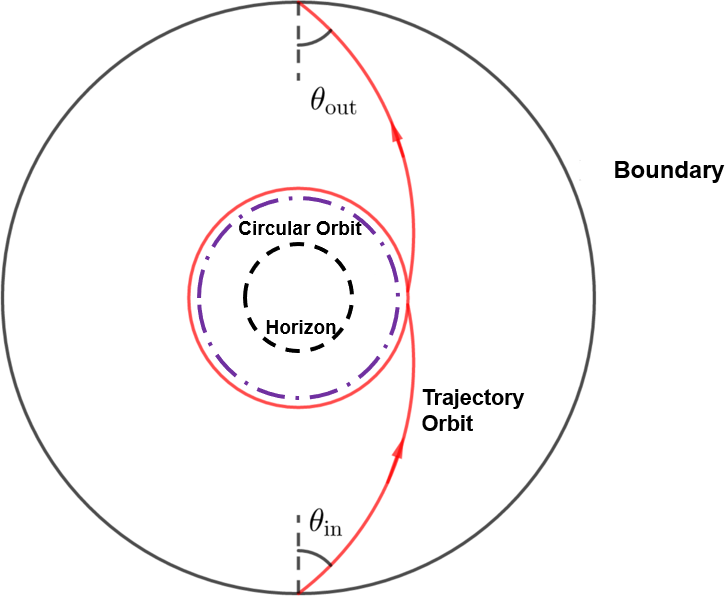}
	\caption{The dominant contribution to the final response function comes from the trajectory as close to the circular orbit as possible.}\label{inp}
\end{figure}

Although $L$ is an integral constant of motion subject to our free choice for a given  large $\omega$,  as illustrated in Figure \ref{inp}, the dominant contribution to the final response function we are considering is supposed to come from the special $L_s$ with which the trajectory emanating from the south pole on the AdS boundary can enter the circular orbit\cite{HKM,HKM2}. This observation also conforms with the relationship between the the circular orbit and  the pole of the aforementioned retarded Green function, or equivalently the quasi-normal mode in the large frequency limit\cite{ZTWS}. Note that this circular orbit is not the trajectory taken by a photon, but nevertheless, we still like to call it  the photon sphere for simplicity. With this in mind, we expect to see a Einstein ring formed on the screen with the ring radius given by
\begin{equation}
   \frac{ r_R}{f}=\frac{L_s}{\tilde{\omega}},
\end{equation}
where $L_s$ together with the circular orbit radius can be determined by the following conditions
\begin{equation}
    \mathcal{R}=0, \quad \frac{d\mathcal{R}}{dr}=0
\end{equation}
for a chosen large $\omega$.

In particular, the value of $\frac{r_R}{f}$ of the Einstein ring formed in Figure \ref{sharpimage} for the high frequency $\omega=75$ is around $0.984$, which is in good agreement with $0.996$, the value given by our geometric optics approximation.

With the above preparation, we shall present our main numerical results in the subsequent section for $\omega=75$, $\sigma=0.05$, and $d=0.6$.

\section{Relevant numerical results}\label{main}
\begin{figure}[t]
    \centering
    \subfigure[$\theta_{\text{obs}}=0$]{
        \includegraphics[width=1.3in]{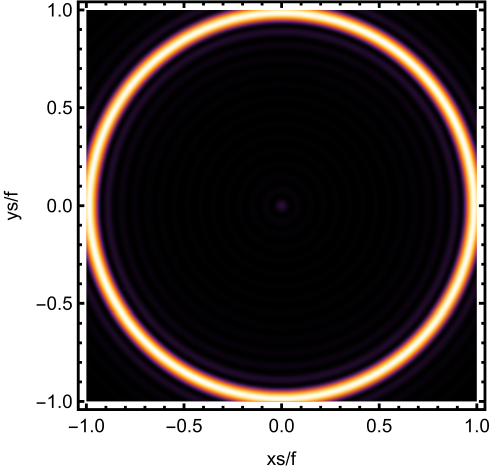}
    }
	\subfigure[$\theta_{\text{obs}}=\pi/6$]{
        \includegraphics[width=1.3in]{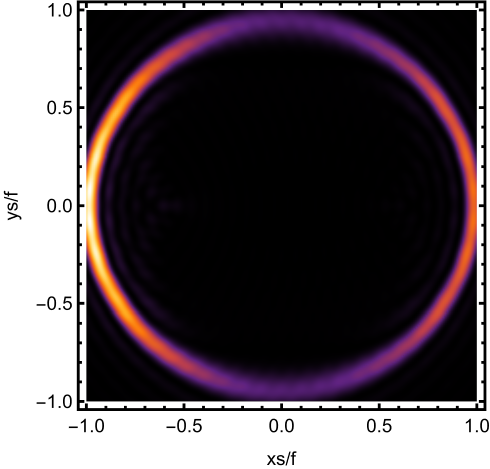}
    }
    \subfigure[$\theta_{\text{obs}}=\pi/3$]{
        \includegraphics[width=1.3in]{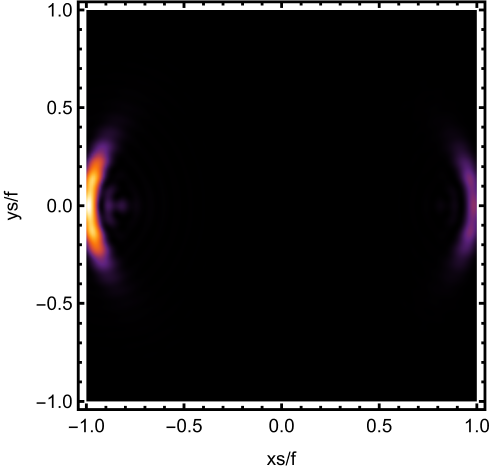}
    }
    \subfigure[$\theta_{\text{obs}}=\pi/2$]{
        \includegraphics[width=1.3in]{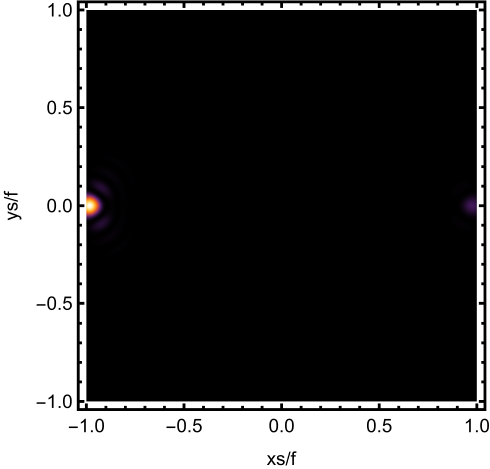}
    }
    \caption{The images of the lensed response observed at various observation angles for $T=1/\pi$ and $\mu=2$.}
    \label{rings1}
\end{figure}

We first plot the typical images of the lensed response observed from different observation angles in Figure \ref{rings1}. With the increase of the observation angle from $\theta_{obs}=0$ to $\theta_{obs}=\frac{\pi}{2}$, the axi-symmetry of the image gets broken gradually from a perfect ring to a bright spot on the left side of the screen. In spite of this, the distance of the image from the center keeps almost unchanged. This is reasonable because as we explain in the previous section, such a distance is determined intrinsically by the circular orbit parameter $L_s$, which
does not vary with the observation angle at all.

\begin{figure}[t]
    \centering
    \subfigure[$T=0.012$]{
        \includegraphics[width=1.3in]{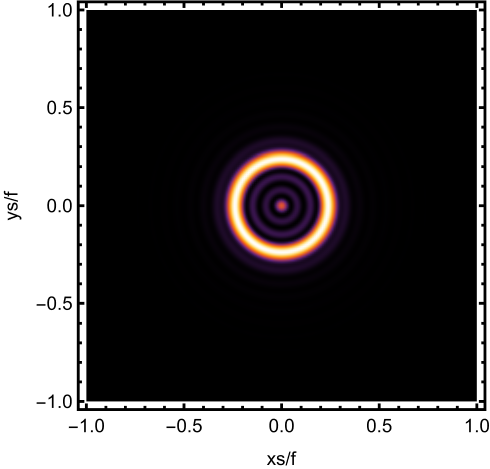}\label{SCHWring}
    }
	\subfigure[$T=0.060$]{
        \includegraphics[width=1.3in]{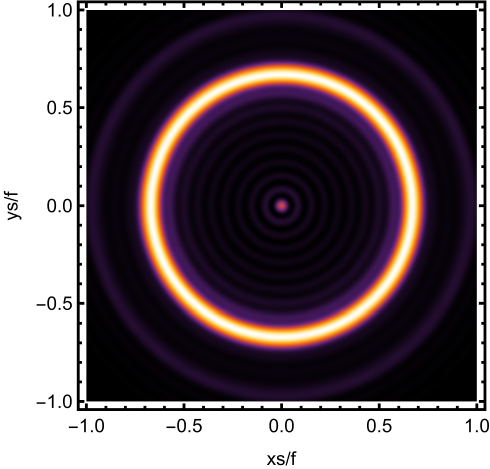}
    }
    \subfigure[$T=0.239$]{
        \includegraphics[width=1.3in]{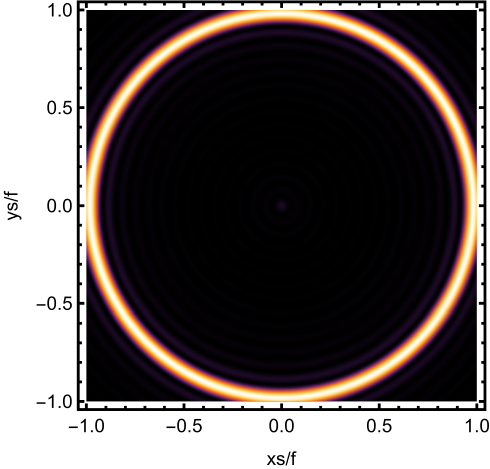}
    }
    \caption{The variation of the image of the lensed response with the temperature at the observation angle $\theta_{obs}=0$ for the fixed chemical potential $\mu=2$. }
    \label{ringstem}
\end{figure}

\begin{figure}[t]
    \centering

    \subfigure[$\mu=0.01$]{
        \includegraphics[width=1.3in]{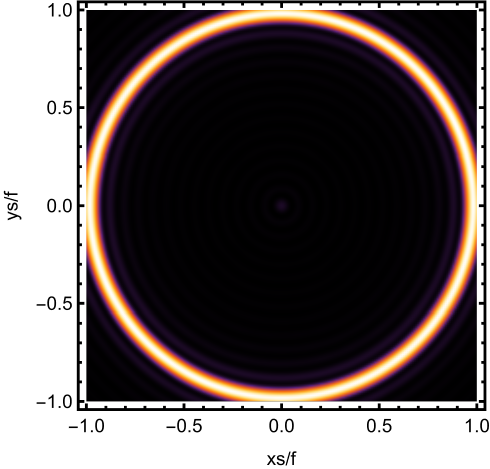}
    }
     \subfigure[$\mu=0.75$]{
        \includegraphics[width=1.3in]{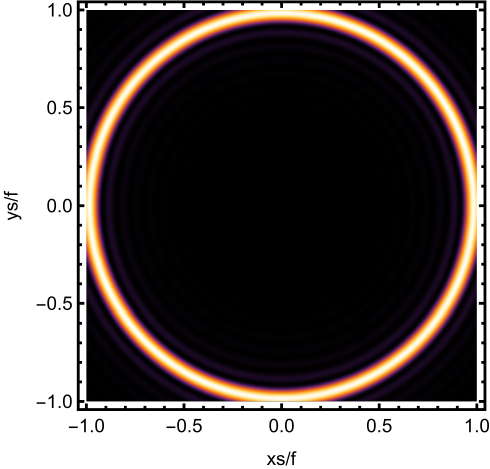}
    }
     \subfigure[$\mu=1.5$]{
        \includegraphics[width=1.3in]{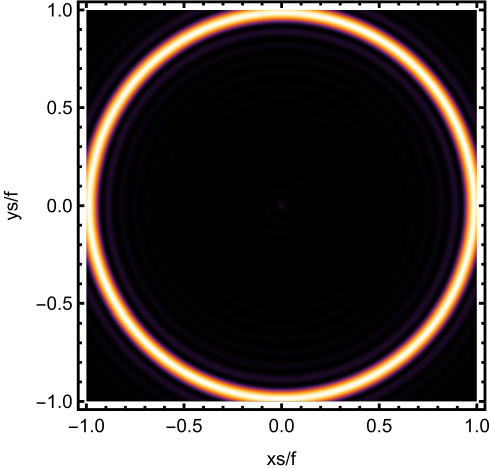}
    }
    \caption{The variation of the image with the chemical potential at the observation angle $\theta_{obs}=0$ for the fixed temperature $T=1/\pi$.}
    \label{ringsdif}
    \end{figure}

Next we plot the variation of the image observed at the north pole  with the temperature in Figure \ref{ringstem}. As we see, not only does the formed Einstein ring lie well within the screen at low temperatures, but also its radius increases with the increase of the temperature in an appreciable manner, as opposed to the aforementioned behavior for the weakly interacting quantum field theory. On the other hand, the images observed at the north pole at different chemical potentials are displayed in Figure \ref{ringsdif}, where we see the variation of the image with respect to the chemical potential is almost negligible, which is similar to the behavior for the weakly interacting quantum field theory.

\begin{figure}
    \centering
    \subfigure[]{
        \includegraphics[width=2.1in]{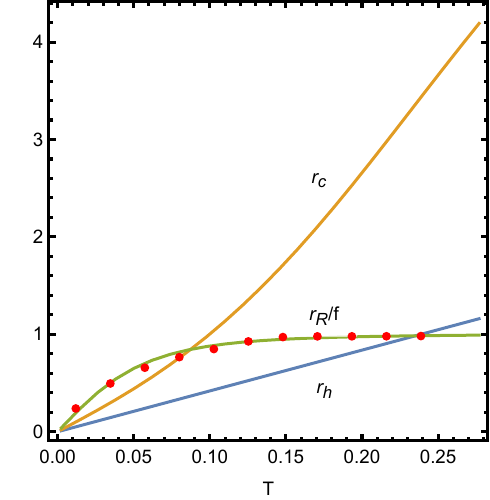}
    }
	\subfigure[]{
        \includegraphics[width=2in]{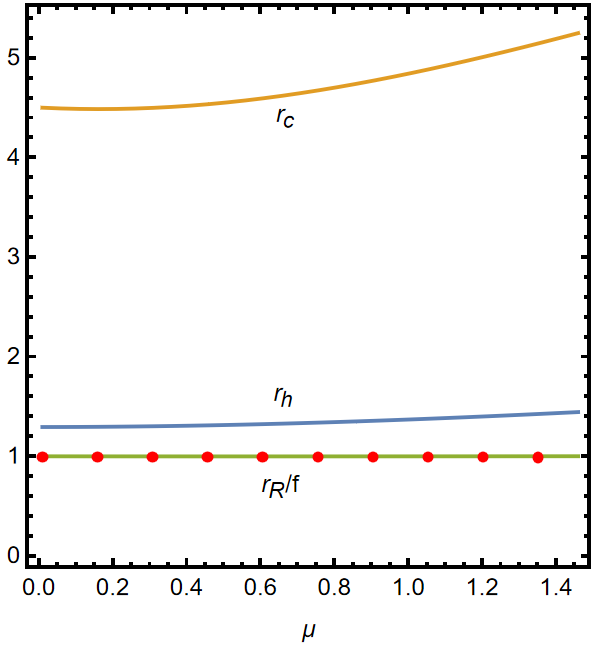}
    }
    \caption{The circular orbit $r_c$, event horizon $r_h$, and ring radius $r_R$ in the unit of $f$ as functions of temperature $T$ at the fixed chemical potential $\mu=2$ in (a) and as functions of chemical potential $\mu$ at the fixed temperature $T=1/\pi$ in (b). The solid curves represent the results obtained by the geometric optics, while the wave optics results are indicated by the discrete red dots.}\label{last}
\end{figure}

We further draw the radius of the Einstein ring in the unit of $f$ as a function of temperature and chemical potential respectively in (a) and (b) of Figure \ref{last}, where both radii of the  black hole horizon and the circular orbit as functions of temperature and chemical potential are also exhibited simply for curiosity's sake. As one can see from (a), the aforementioned appreciable increase in the Einstein ring radius occurs only at low temperatures.  After this increase, the Einstein ring radius starts to flatten out. On the other hand, (b) tells us that the Einstein ring radius keeps unchanged indeed as one cranks up the chemical potential in a continuous manner. Last, as expected, the Einstein ring radius obtained by our wave optics fits well with that by geometric optics.

\section{Conclusion}\label{conclusion}
Motivated by the fact that the real quantum materials are engineered generically at a finite chemical potential, we have investigated the resulting Einstein ring structure for the lensed response of the complex scalar field as a probe wave propagating in the RN-AdS black hole in the context of AdS/CFT. Among others, we like to highlight one similarity and one distinction we have found between the weakly interacting quantum system and the strongly coupled one to which our RN-AdS black hole is dual. The similarity is that the ring radius keeps unchanged with the increase of the chemical potential for both systems. On the other hand, the distinction lies in the fact that not only can the Einstein ring exist well within the screen, but also the radius exhibits an appreciable increase at low temperatures for the holographic system while the ring radius for the weakly interacting system displays no temperature dependence, still keeping unchanged right at the edge of the screen. Note that our holographic Einstein ring emerges in the large frequencies, which has been shown to be related to the bulk generalized photon sphere away from the black hole horizon in the geometric optics approximation. Thus such a distinct behavior may be regarded as a universal feature associated with the high energy modes of the holographic system. With this in mind, one can use this feature to diagnose whether the quantum system under consideration is strongly coupled to have a gravity dual description or not.

We conclude our paper with some future directions. First, what we have explored so far is only for the response function of the scalar operator, so it is interesting to see what happens to the response function of the fermionic operator, which is supposed to shed new light into the characteristic features for non-Fermi liquids\cite{FILMV,LMV}.
In addition, as alluded to in the introduction section, the holographic gravity dual we consider is the simplest one for the boundary thermal system at finite chemical potential. There are a variety of more sophisticated holographic models in action. It is intriguing to investigate the detailed behavior of the lensed response function for such models as the holographic superconductor, which has been partially studied in \cite{KMT} indeed. Last but not least, the underlying numerics we have developed provides us with  an efficient route to image the black hole by wave optics. Thus it is worthwhile to employ our numerical techniques to reconstruct the images of the real life black holes in the centers of our galaxy and nearby galaxies. Compared to the conventional observer-oriented ray tracing method by geometric optics, our wave optics is source-based, which may has its advantages in some situations, if not all. Moreover, it is supposed to demonstrate us richer information about the black holes in the sky since geometric optics is only the approximation of wave optics after all.

\appendix

\section{Retarded Green function in thermal field theory at a finite chemical potential}\label{AA}

\begin{figure}
    \centering
    \subfigure[]{
        \includegraphics[width=2.1in]{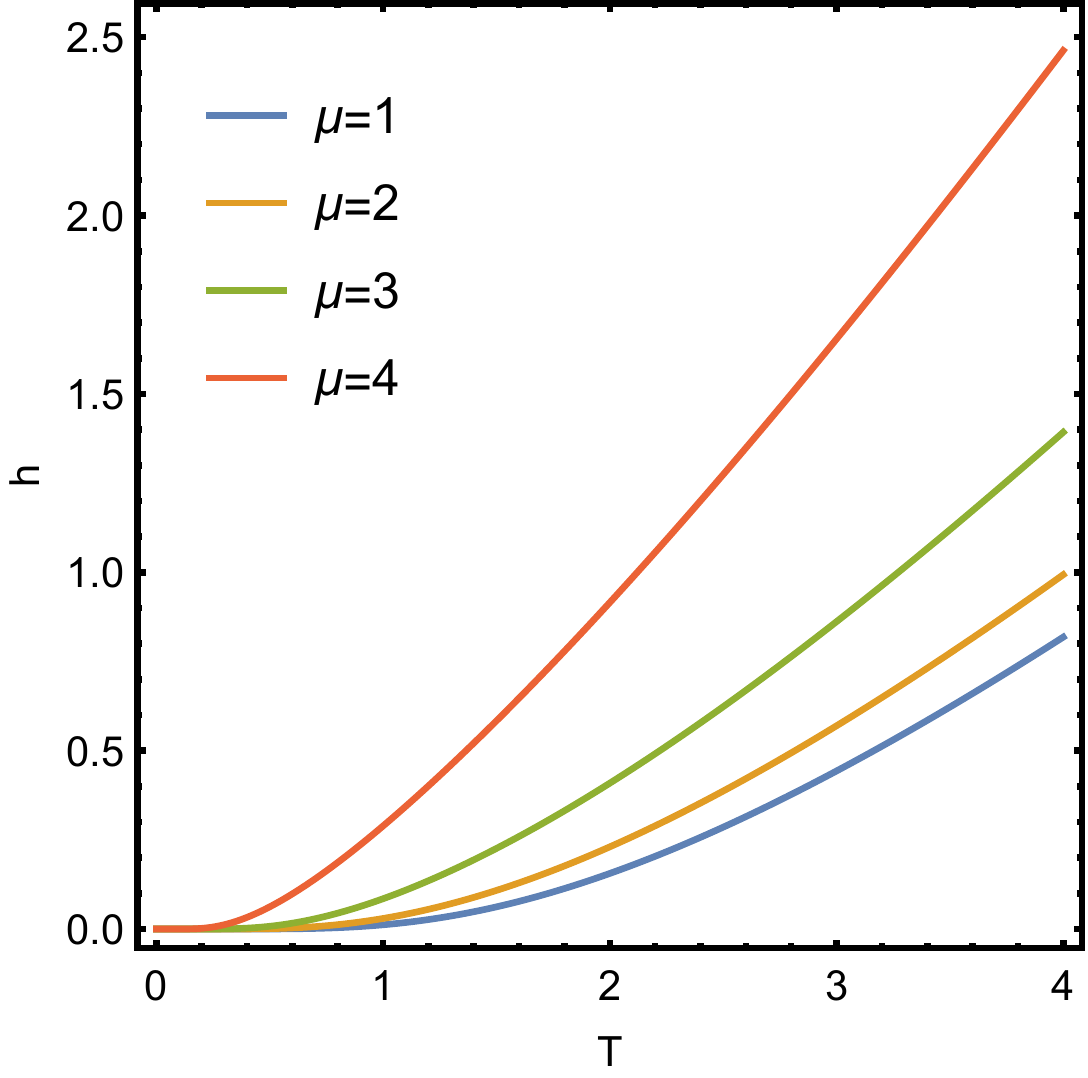}
    }
	\subfigure[]{
        \includegraphics[width=2.1in]{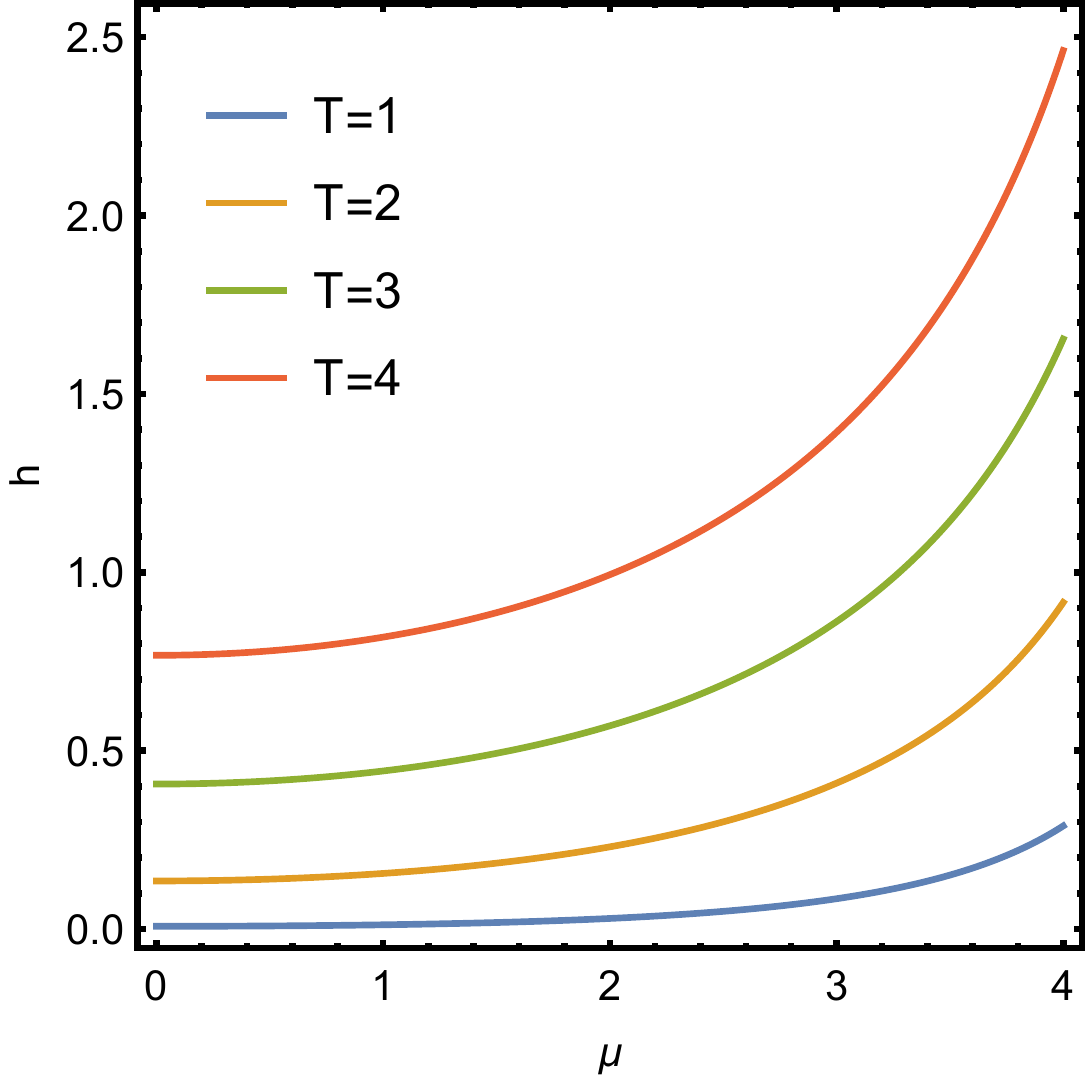}
    }
    \caption{The dependence of the correction term on the temperature and the chemical potential for $m=5$.}\label{weakly}
\end{figure}

Let us consider the simplest Lagrangian with a global U(1) symmetry
\begin{equation}
    \mathcal{L}=-\nabla_a\bar{\phi}\nabla^a\phi-m^2\bar{\phi}\phi-V(\bar{\phi}\phi)
\end{equation}
under the transformation $\phi\rightarrow e^{-i\alpha}\phi$ in the following spacetime \begin{equation}
    \exd s^2=-\exd t^2+h_{ij}\exd x^i\exd x^j.
\end{equation}
The corresponding Hamiltonian and conserved charge are given by
\begin{equation}
H=\int \exd\mathbf{x}\sqrt{h}(\bar{\pi}\pi+h^{ab}\nabla_a\bar{\phi}\nabla_b\phi+m^2\bar{\phi}\phi+V),\quad
Q=i\int \exd\mathbf{x}\sqrt{h}(\bar{\phi}\pi-\phi\bar{\pi})
\end{equation}
with $\pi=\partial_t\phi$ and $\bar{\pi}=\partial_t\bar{\phi}$.  Accordingly, we have
\begin{eqnarray}
   \bra\phi'| e^{-i(H-\mu Q)t}|\phi\ket&=&\int_\phi^{\phi'}D\phi\int D\pi e^{i\int_0^t dt\int d\mathbf{x}\sqrt{h} [\bar{\pi}\partial_t\phi+\pi\partial_t\bar{\phi}+i\mu(\bar{\phi}\pi-\phi\bar{\pi})-\bar{\pi}\pi-h^{ab}\nabla_a\bar{\phi}\nabla_b\phi-m^2\bar{\phi}\phi-V]}\nonumber\\
  &=& \int_\phi^{\phi'}D\phi\int D\pi' e^{i\int_0^t dt\int d\mathbf{x}\sqrt{h} [-\bar{\pi}'\pi'+(\partial_t+i\mu)\bar{\phi}(\partial_t-i\mu)\phi-h^{ab}\nabla_a\bar{\phi}\nabla_b\phi-m^2\bar{\phi}\phi-V]}\nonumber\\
  &=&C\int_\phi^{\phi'}D\phi\int e^{i\int_0^t dt\int d\mathbf{x}\sqrt{h} [(\partial_t+i\mu)\bar{\phi}(\partial_t-i\mu)\phi-h^{ab}\nabla_a\bar{\phi}\nabla_b\phi-m^2\bar{\phi}\phi-V]},
\end{eqnarray}
where $\pi'=\pi-\partial_t\phi+i\mu\phi$ and $\bar{\pi}'= \bar{\pi}-\partial_t\bar{\phi}-i\mu\bar{\phi}$. With $it=\tau$, the partition function in the grand canonical ensemble can be written in terms of the following path integral
\begin{eqnarray}
   \mathcal{Z}(T,\mu)&\equiv&\text{Tr}[e^{-\beta(H-\mu Q)}]\nonumber\\
        &=&C\int D\phi e^{-\int_0^\beta d\tau \int d\mathbf{x}\sqrt{h}[(\partial_\tau+\mu)\bar{\phi}(\partial_\tau-\mu)\phi+h^{ab}\nabla_a\bar{\phi}\nabla_b\phi+m^2\bar{\phi}\phi+V]}\nonumber\\
                &=&C\int D\phi e^{-\int_0^\beta d\tau \int d\mathbf{x}\sqrt{h}\{\bar{\phi}[-(\partial_\tau-\mu)^2-\tilde{D}_a\tilde{D}^a+m^2]\phi+V\}},
        \end{eqnarray}
        where the periodic boundary condition is assumed along the $\tau$ direction with $T=\frac{1}{\beta}$ and $\tilde{D}_a$ the induced spatial covariant derivative.
        Whence the thermal Green function for the free field theory reads
        \begin{equation}
        G^0_E(x,x')=\bra x|\frac{1}{-(\partial_\tau-\mu)^2-\tilde{D}_a\tilde{D}^a+m^2}|x'\ket.
        \end{equation}
        When the spatial metric is the $2$-sphere, $\tilde{D}_a\tilde{D}^a$ is simply $D_S^2$ and the resulting free thermal Green function can be expressed as
        \begin{equation}
            G^0_E(x,x')=\frac{1}{\beta}\sum_{n,l,m}\frac{1}{-(-i\omega_n-\mu)^2+l(l+1)+m^2}e^{-i\omega_n(\tau-\tau')}Y_{lm}(\theta,\varphi)\bar{Y}_{lm}(\theta',\varphi'),
        \end{equation}
where $\omega_n=\frac{2n\pi}{\beta}$ are called the Matsubara frequencies. By the usual analytic continuation, one ends up with the retarded Green function as
\begin{equation}
    G^0_{lm}(\omega)=-G^0_{Elm}[-i(\omega+i\epsilon)]=\frac{1}{(\omega+\mu)^2-l(l+1)-m^2}.
\end{equation}
For $V=\frac{\lambda}{4}(\bar{\phi}\phi)^2$ with $\lambda$ the weak coupling parameter, the self-energy at one-loop level is given by
\begin{eqnarray}
    \Pi&=&-\frac{\lambda}{\beta}\sum_{n,l}\frac{1}{(\omega_n-i\mu)^2+l(l+1)+m^2}\nonumber\\
    &=&-\frac{\lambda}{\beta}\sum_{\omega\in \frac{2\pi n}{\beta},l}\frac{\beta}{2}\text{Res}[\cot\frac{\beta\omega}{2}\frac{1}{(\omega-i\mu)^2+l(l+1)+m^2}]\nonumber\\
    &=&\frac{\lambda}{2}\sum_{\omega\notin\frac{2\pi n}{\beta},l}\text{Res}[\cot\frac{\beta\omega}{2}\frac{1}{(\omega-i\mu)^2+l(l+1)+m^2}]\nonumber\\
    &=&-\frac{\lambda}{4}\sum_l\frac{1}{\omega_l}[\coth\frac{\beta(\omega_l+\mu)}{2}+\coth\frac{\beta(\omega_l-\mu)}{2}]\nonumber\\
    &=&-\frac{\lambda}{2}\sum_l\frac{1}{\omega_l}[1+\frac{1}{e^{\beta(\omega_l+\mu)}-1}+\frac{1}{e^{\beta(\omega_l-\mu)}-1}]
\end{eqnarray}
with $\omega_l=\sqrt{l(l+1)+m^2}$. The first term, coming from the pure vacuum one-loop contribution, is divergent and can be renormalized to zero. The rest two terms, corresponding to the corrections induced by the finite temperature and finite chemical potential, turns out to be finite and  will correct the retarded Green function at the finite temperature and finite chemical potential through
\begin{equation}
m^2\rightarrow m^2+\frac{\lambda}{2}h(T,\mu)
\end{equation}
with
\begin{equation}
    h=\sum_l\frac{1}{\omega_l}[\frac{1}{e^{\beta(\omega_l+\mu)}-1}+\frac{1}{e^{\beta(\omega_l-\mu)}-1}].
\end{equation}
However, this correction term is not amenable to an analytic treatment. So we resort to the numerics to show its typical dependence on the temperature and the chemical potential in Figure \ref{weakly}. As we see, for the fixed chemical potential, the variation of $h$ is negligible at low temperatures. In addition, the larger the fixed chemical potential becomes, the lower the threshold temperature for $h$ to start growing in an almost linear manner becomes. On the other hand, for the fixed temperature, the non-negligible variation of $h$ occurs from some large chemical potential. Moreover, the higher the temperature becomes, the smaller such a threshold chemical potential becomes.
\section{Response extractions by pseudo-spectral method}\label{spectral}
The key task in obtaining the response is to solve
$Z_l(z)$ by the bulk equation (\ref{zpart}), which can be achieved by the pseudo-spectral method. For the finite computational domain $z\in(0,z_h)$, we expand $Z_l$ as

\begin{equation}
	Z_l(z)=\sum_{i=0}^N a_i\tilde T_i(z),
\end{equation}
where $\tilde T_i(z)\notice T_i(2z/z_h-1)$ with $T_i(\cos\theta)=\cos(i\theta)$ the Chebyshev polynomials.
Next with the choice of
$N+1$ collocation points $z_j$ as
\begin{equation}
	z_j=\frac12(z_hx_j+z_h)=\frac12\left[z_h\cos\frac{\pi j}{N}+z_h\right], j=0,1,\cdots,N,
\end{equation}
the differential with respect to $z$ can be discretized as the following differential matrix
\begin{equation}
	\mbf{\tilde{D}_N}=\frac{2}{z_h}\mbf{D_N},
\end{equation}
where $\mbf{D_N}$ can be expressed explicitly as

\begin{equation}
 	\mbf{D} _{\mbf{N}00}=\frac{2N^2+1}{6},\qquad
 	\mbf{D}_{\mbf{N}NN}=-\frac{2N^2+1}{6},
\end{equation}
\begin{equation}
	\mbf{D} _{\mbf{N}jj}=\frac{-x_j}{2(1-x_j^2)},\qquad j\neq0,N,
\end{equation}
\begin{equation}
	\bm{D} _{\mbf{N}ij}=\frac{c_i}{c_j}\frac{(-1)^{i+j}}{x_i-x_j},\qquad i\neq j,
\end{equation}
with $c_i$ the coefficients such that
\begin{equation}
	c_i=\begin{cases}
	2\quad \text{if}\ i=0\ \text{or}\ N,\\
	1\quad \text{otherwise}.
	\end{cases}
\end{equation}
Accordingly, the bulk differential equation (\ref{zpart}) is transformed into an algebraic equation, which together with the boundary condition $Z_l(0)=1$ can be solved readily by numerics. Finally, the response can be extracted according to the following formula
\begin{equation}
	\bra O\ket_l=\left(\mbf{\tilde{D}_N}\right)_{Ni}Z_l(z_i).
\end{equation}

\section*{Acknowledgements}{LYX is grateful to Yu-Chen Ding and Qing-Hua Zhu for their helpful discussions.  He also thanks his wife for her supporting his work on the honeymoon.
This work is supported in part by the National Natural Science Foundation of China with Grant No. 11875095 and 12075026, as well as by China Postdoctoral Science Foundation, under the National Postdoctoral Program for Innovative Talents BX2021303.
}

%\bibliography{a}

\end{document}